# Consequences of Gravitational Tunneling Radiation

Mario Rabinowitz*

**Abstract**

Gravitational tunneling radiation (GTR) from a black hole is similar to electric field emission of electrons from a metal. There is always a second body to give the gravitational barrier of a black hole a finite rather than infinite width so that a particle can escape by tunneling. Thus GTR may be emitted from black holes in a process differing from that of Hawking radiation (HR) with broad ranging astrophysical ramifications. HR of an "isolated" black hole is an abstraction which has not been detected in three decades, and which may not exist. GTR is beamed, in contrast to HR, which is isotropic. Little black holes may account for much of the dark matter of the universe, and hence the anomalous rotation of spiral galaxies, as well as contribute to the accelerated expansion of the universe. The absence of HR permits the existence of little black hole gravitationally bound atoms with quantized orbits and quantized gravitational radiation. Since some theories speculate that the early universe may have existed in n-dimensional space, certain results will be presented both for general n-space and for the 3-space in which we live. An inconsistency between quantum mechanics and the weak equivalence principle is demonstrated. It is shown that angular momentum cannot be quantized in the usual manner in 4-space, with possible consequences for Kaluza-Klein theory. An hypothesis is presented for reducing the enormously too-large vacuum zero-point energy. String theory may be impacted by the finding that atoms are not energetically stable in higher than 3-space. GTR avoids the black hole lost information paradox.

# 1  Introduction

Since dark matter, the missing mass of the universe, is ~ 20 times more plentiful than visible matter, it may have played a key role in shaping the universe we see. Small density fluctuations in the early universe are thought to be the gravitational origin of the structure of the present universe. If so, fluctuations and ensuing motion of the dark matter may have played a crucial role in the dynamics and morphology of the universe in which galaxies are found in hierarchies of clusters and superclusters separated by immense voids. Among other things, this paper explores the hypothesis that if little black holes (LBH) radiate by gravitational tunneling radiation (a process that differs substantially from Hawking radiation), some of the not well understood features of the universe can be accounted for.

# 2  Gravitational Tunneling Radiation

Gravitational tunneling radiation (GTR) originates from within a black hole and tunnels out due to the field of a second body (in contrast to Hawking's single body approach), which gives the barrier a finite width. Even though general relativity does not deal with potentials, this is schematically depicted in Fig. 1 for the purpose of an intuitive picture. GTR is similar to electric field emission of electrons from a metal by the application of an external field, except that a replenishing source of mass is not needed since there is only one sign of mass.  In field emission, electrons must be replenished at the emitter since it becomes less negative as electrons leave it.

*Armor Research, 715 Lakemead Way, Redwood City, CA 94062-3922; Mario715@earthlink.net

GTR can be emitted at much higher intensities and temperatures in the present epoch than HR because smaller LBH can still exist as will be shown in Sec. 6.  GTR also sheds light on why HR has not yet been observed,  Thorne et al (1986) have calculated that LBH emission is ~ 37 %  by electrons and positrons, ~ 8 % by photons, and the remainder  equally divided by the six kinds of neutrinos together with a small component of higher mass particles and gravitons. Argyres et al (1998) conclude that the properties of LBH are greatly altered and LBH radiation is considerably attenuated from that of Hawking's prediction.  Their LBH are trapped by branes so that essentially only gravitons can get through the brane (which may be thought of as an abbreviation for vibrating membrane).  For them, not only is the radiation rate as much as a factor of $10^{38}$ lower, but it also differs by being almost entirely gravitons.

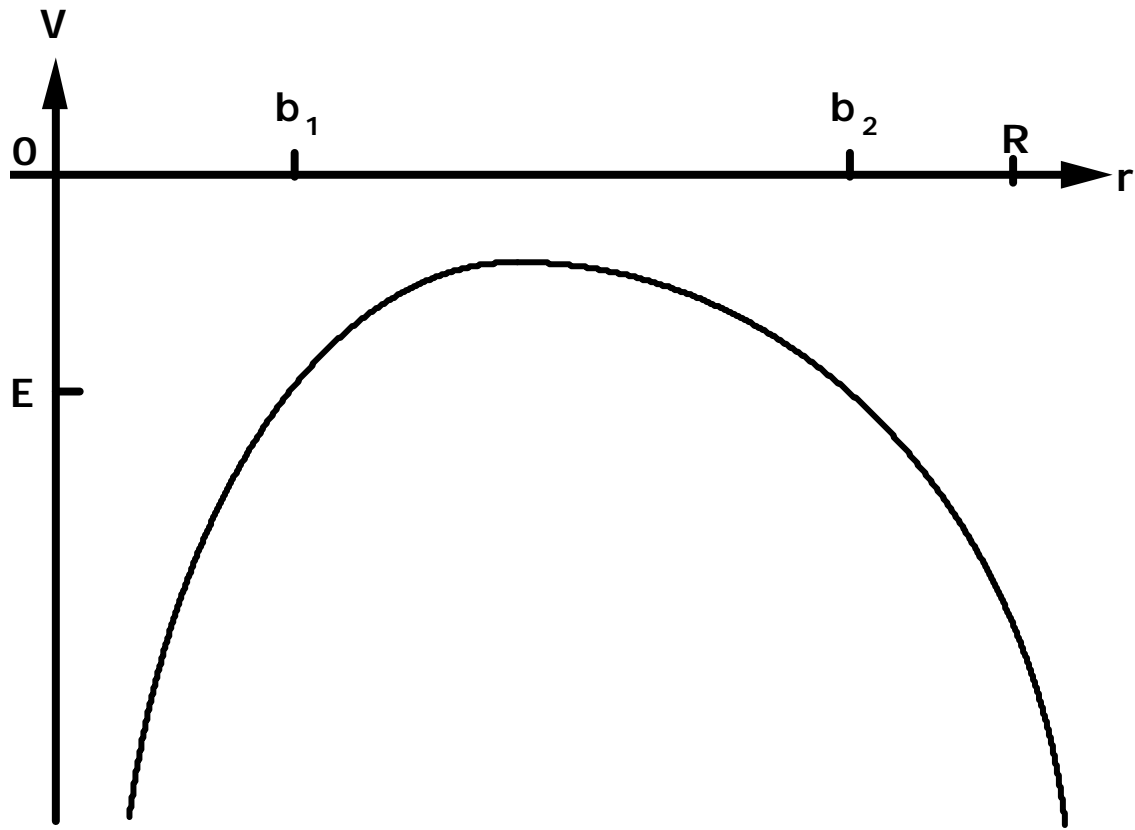

**Fig. 1.**  Heuristic representation of the gravitational barrier with classical turning points $b_1$ and $b_2$, and a second body at R, opposite a black hole at the origin.

Gravitational tunneling radiated power, $P_R$, that tunnels out of a LBH is:

$$P_R \approx \left|\frac{\hbar c^6 \langle e^{-2\Delta\gamma}\rangle}{16\pi G^2}\right| \frac{1}{M^2} = \frac{\langle e^{-2\Delta\gamma}\rangle}{M^2}\left[3.42 \times 10^{35} W\right], \tag{2.1}$$

where M in kg is the mass of the LBH (Rabinowitz,1999 a, b, c). Although $P_R$ is of a different physical origin than Hawking radiation, it is analytically of the same form since $P_R \propto \Gamma P_{SH}$, where $\Gamma$ is the transmission probability approximately equal to the WKBJ tunneling probability $e^{-2\Delta\gamma}$ for LBH. The tunneling probability $e^{-2\Delta\gamma}$ is usually << 1 and depends on parameters such as the width of the barrier, M, and the mass of the second body.

Two LBH may get quite close for maximum tunneling radiation. In this limit, there is a similarity between the tunneling model, and what is expected from the Hawking model (1974,1975) in that the tidal forces of two LBH add together to give more radiation at their interface in HR, also producing a repulsive force.

GTR emission is greatest when the companion is a nearby LBH, and least when it is a distant ordinary body. The escaping particle may be trapped in the well of the second body. If it is not also a black hole, then escape from it can occur by ordinary processes such as scattering and gravity-assisted energy from the second body's angular momentum.

A distinction needs to be made between the concepts of transmission and tunneling as used in the terms "transmission probability or transmission coefficient " and "tunneling probability or penetration coefficient." Tunneling probability, $\Pi$, is a ratio of probability densities, and transmission probability, $\Gamma$, is a ratio of probability current densities. For an Einsteinian black hole there can be a relatively small barrier at 1.5 $R_H$ that even virtual particles have to tunnel through to get out. Hawking (1975) uses the terms "tunnel" or "penetrate" through this potential barrier interchangeably, and does not use the term "transmission." Since his calculations or final results for these quantities are not presented, it appears from context that he is using both these terms just for "tunneling probability." The much earlier literature also often did not distinguish between the two concepts. The transmission probability or coefficient

$$\Gamma \equiv \frac{\Psi(b_2)\Psi^*(b_2)v_2}{\Psi(b_1)\Psi^*(b_1)v_1} \approx \frac{v_2}{v_1} e^{-2[\gamma(b_2)-\gamma(b_1)]} = \Pi\frac{v_2}{v_1} , \tag{2.2}$$

where (cf. Fig.1 as a heuristic aid to visualization) tunneling is from region 1 (left of the barrier, 0 to classical turning point $b_1$) to region 2 (right of the barrier from classical turning point $b_2$ to R, the location of the second body $M_2$). $\Gamma = \Pi$ when the velocities $v_1$ and $v_2$ are the same on both sides of the barrier.

In general, without needing an explicit solution for the wave function $\Psi$, it can be shown (Rabinowitz, 1999a) that

$$\Gamma = \frac{\Psi_2 \Psi_2^*}{\Psi_{inc} \Psi_{inc}^*}\left(\frac{v_2}{v_1}\right) = \left[e^{\Delta\gamma} - \tfrac{1}{4} e^{-\Delta\gamma}\right]^{-2} . \tag{2.3}$$

where

$$\Delta\gamma \equiv \int_{b_1}^{b_2}\left[\frac{2m}{\hbar^2}(V-E)\right]^{1/2} dr. \tag{2.4}$$

Thus when $\Delta\gamma$ is large, $e^{\Delta\gamma} \gg (1/4)e^{-\Delta\gamma}$ in eq. (2.3), yielding $\Gamma \approx \Pi = e^{-2\Delta\gamma}$. $\Gamma \approx \Pi$ is true in most cases when $b_2 - b_1 \gg 0$, and/or $V \gg E$.

However, we shall be mainly interested in the high energy case, when V - E is small which (for our gravitational barriers) implies that the distance between the classical turning points, $b_2 - b_1$ may not be relatively large. At first sight it would appear that we cannot make the approximation $\Gamma \approx \Pi$. Propitiously, the barrier of Fig. 1 becomes symmetrical for all energies and barrier widths when $M = M_2$ and then $v_1 = v_2$. Similarly $v_1 \sim v_2$ for $M \sim M_2$. So $\Gamma \approx \Pi$ is a valid approximation when $M \sim M_2$, and is quite accurate for all M and $M_2$ in the case of ultrarelativistic particles, where $v_1 ♠ v_2 ♠ c$, the speed of light. However, for non-zero rest mass particles, when their energies are low in a non-symmetrical gravitational barrier, this may not be a valid approximation. This seems to have been overlooked by Hawking and others. It is a good approximation for little black holes because low energy particles are a miniscule fraction of the radiation due to extremely high temperatures. There is so little radiation from high mass black holes, that this correction can usually be neglected. However it needs to be taken into consideration for intermediate mass black holes.

## 3 Black Hole Temperature

The Hawking 1974 value for temperature is a factor of 2 smaller than his 1975 value. This is not critical, and the 1975 expression is

$$T = \left[\frac{\hbar c^3}{4\pi kG}\right]\frac{1}{M} = \left[2.46 \times 10^{23}\right]\left(\frac{1}{M}\right) {}^{o}K , \tag{3.1}$$

with M in kg. For $M \sim 10^{12}$ kg (the smallest primordial black hole mass that can survive to the present for Hawking), $T \sim 10^{11}$ K. As we shall see in Sec. 6, GTR permits the survival of much smaller masses such as for example $M \sim 10^{-3}$ kg with $T \sim 10^{26}$ K.

Particles that originate at or outside the horizon of an isolated black hole must lose energy in escaping the gravitational potential of the black hole. For example, in the case of photons

$$\nu_{observer} = \nu(r_{emission\ site})\left[1 - \frac{2GM/c^2}{r}\right]^{1/2} = \nu(r_{emiss\ site})\left[1 - \frac{R_H}{r}\right]^{1/2}, \quad (3.2)$$

where $\nu_{observer}$ is the frequency detected by the observer at a very large distance from the black hole, and $\nu(r_{emission\ site})$ is the frequency at the radial distance r from the center of the black hole where the particle pair was created outside the black hole. Eq. (3.2) does not depend on the gravitational potential between the emission site and observer, but only on the emission proximity to $R_H$. We see that for $r = R_H$ and any finite $\nu(r_{emission\ site})$, $\nu_{observer} = 0$ implying that any finite temperature at $R_H$ must red shift to zero as measured at large distances from a black hole.

Thus the Hawking temperature appears to have an inconsistency. Although originally proposed as not real by Bardeen, Carter, and Hawking (1973), this temperature is now asserted and generally accepted as the gravitationally red shifted temperature. As shown by eq. (3.2), their new view implies infinite temperature and infinite radiation frequency at the horizon of all black holes, since the red shift goes to zero as measured at large distances from any hole if the surface temperature or frequency are finite. For the real temperature, they averred "the effective temperature of a black hole is zero ... because the time dilation factor [red shift] tends to zero on the horizon." Temperature could be inferred for an LBH from the energy distribution of emitted particles. In GTR there is relatively little red shift, as the particles tunnel through the barrier with undiminished energy. What little red shift there is, starts far from the black hole horizon.

As can be seen from eq. (3.1), the black hole temperature is inversely proportional to its mass. Gravitational systems tend to exhibit such a negative heat capacity. This peculiarity is not limited to black holes, and occurs in many cases in addition to Einsteinian and Newtonian gravity. Black holes oddly get hotter (rather than cooler) with a local decrease and global increase in entropy, the more energy they lose by radiation. This is ascribable to quantum mechanics because as the hole shrinks, the quantum wavelength must also decrease leading to a higher momentum and hence higher temperature. A classical black hole should not shrink by evaporation because it supposedly cannot radiate. But if it could, it would get cooler. Dunning-Davies and Lavenda (1998) have clearly examined thermodynamic anomalies associated with the Hawking black hole model.

The following examples show that black holes are not the only things that unexpectedly increase their temperature. A non-quantum mechanical

Newtonian example would be satellites losing energy as they move through the atmosphere causing them to gain kinetic energy (at the loss of overall energy and gravitational potential energy) as they fall towards the Earth. This is because they lose angular momentum due to the torque exerted by air resistance. This would also be true in the analogous macroscopic electrostatic case. For a high density ensemble of macro-particles, having a high collision frequency as they fall into a central force as they orbit in a viscous medium, their effective non-equilibrium temperature will increase.

The above satellite example does not apply to satellites upon which the dissipative force exerts no torque. Lunar tidal friction decreases both the Earth's and the moon's energy but hardly, if at all, exerts torque on the moon to decrease its angular momentum. In fact the moon's angular momentum may even increase as the spin momentum of the Earth decreases. Even if the moon's net angular momentum were conserved, tidal dissipation would cause the moon to move further from the Earth with a decrease in kinetic energy.

In a Joule-Thomson expansion, one usually observes a lowering of the temperature because the total energy is conserved, and in most cases the potential energy increases with the expansion leading to a lower kinetic energy and hence lower temperature. However, at high temperature and/or high pressure, expansion can decrease the potential energy, and increase the temperature.

## 4 Two-Body, M>>m, Gravitational Atoms in n-Space

Ordinary matter does not have a high enough density to make gravitationally bound atoms feasible, but LBH do. However, HR would prohibit such atoms from staying together. One consequence of GTR is that LBH radiation may be sufficiently low for a large enough mass LBH and a very small orbiting mass to make such atoms possible. Let us explore this possibility, ignoring for now the effects of radiation. We will consider n-space of the (n + 1) space-time manifold, since some theories attribute physical reality to macroscopic higher dimensions in the early universe, which should abound with primordial LBH.

To create a black hole in 3-space, an object of mass M (kg) must be crushed to a density

$$\rho = m / \left[\tfrac{4}{3} \pi R_H^3\right] = 7.3 \times 10^{79} M^{-2} \text{ kg}/\text{m}^3, \tag{4.1}$$

where the black hole radius

$$R_H = 2GM/c^2 = 1.48 \times 10^{-31} M \text{ meters}. \tag{4.2}$$

So a 1 kg LBH has a radius of only $10^{-31}$ m and an enormous density of $10^{80}$ kg/m$^3$, compared with a proton of mass $10^{-27}$ kg, a radius of $10^{-15}$ m, and a density of $10^{18}$ kg/m$^3$. We will consider 2-body quantized non-relativistic gravitational orbits in Euclidean n-space, where n = 3, 4, 5 ... is the number of spatial dimensions in the space-time manifold of (n+1) dimensions. Such orbits are an analog of electrostatic atomic orbitals. At least one of the bodies is a black hole of mass $M \gg m$, the mass of the orbiting body.

Without going through a complete derivation again, we can use the previously derived semi-classical results for circular orbits (Rabinowitz, 2001a). One might challenge the use of semi-classical physics at such a small scale and high energies. However, as measured at large distances, the gravitational red shift substantially reduces the impact of high energies near LBH. Argyres et al (1998) argue that "...one can describe black holes by semi-classical physics down to much smaller masses of order of the fundamental Planck scale... ," where the Planck mass of $10^{-8}$ kg has a LBH radius of $10^{-35}$ m. (There were some inadvertent typos that were inconsequential in the limit of 3-space. Due to the complexity of the general n-space equations, it is clearer and easier to present the corrected results here than to refer the reader to the original equations with small changes here and there.)

For $M \gg m$, equating the gravitational force and the centripetal force

$$F_n = \frac{-2\pi G_n M m \Gamma\left(\frac{n}{2}\right)}{\pi^{n/2} r^{n-1}} = \frac{-m v_n^2}{r_n}, \quad (4.3)$$

where the n-space universal gravitational constant $G_n$ changes, in a way that is model dependent, from its 3-space value. The Gamma function $\Gamma(n) \equiv \int_0^\infty t^{n-1} e^{-t} dt$ for all n (integer and non-integer). When n is an integer, $\Gamma(n) = (n-1)!$. $\hbar$ is (Planck's constant)/$2\pi$. Eq. (4.3) implies that the n-space orbital velocity is

$$v_n = \left[\frac{2\pi G_n M \Gamma\left(\frac{n}{2}\right)}{\pi^{n/2} r^{n-2}}\right]^{1/2}. \quad (4.4)$$

From the Bohr-Sommerfeld condition, $m v_n r_n = j\hbar$, we find for the orbital radius of m around M

$$r_n = \left[\frac{j\hbar \pi^{\frac{n-2}{4}}}{m[2 G_n M \Gamma(n/2)]^{1/2}}\right]^{\frac{2}{4-n}}. \quad (4.5)$$

In 3-space eq. (4.5) yields $r_3 = j^2 \hbar^2 / G M m^2$.

Substituting eq. (4.5) into eq. (4.4), the orbital velocity is

$$v_n = \left\{ \left[ \frac{2\pi G_n M \Gamma\left(\frac{n}{2}\right)}{\pi^{n/2}} \right] \left[ \frac{m^{2/(4-n)} \left[ 2G_n M \Gamma\left(\frac{n}{2}\right) \right]^{\frac{1}{4-n}}}{(j\hbar)^{2/(4-n)} \pi^{(n-2)/2(4-n)}} \right] \right\}^{1/2} \tag{4.6}$$

In 3-dimensions eq. (4.6) gives $v_3 = GMm / j\hbar$.

Using eqs. (4.3) and (4.5), the acceleration of the orbiting mass m is

$$a_n = \frac{F_n}{m} = \frac{-2\pi G_n M \Gamma\left(\frac{n}{2}\right)}{\pi^{n/2}} \left[ \frac{m[2G_n M \Gamma(n/2)]^{1/2}}{j\hbar \pi^{(n-2)/4}} \right]^{\frac{2n-2}{4-n}}. \tag{4.7}$$

In 3-dimensional space, eq. (4.7) yields $a_3 = -G^3 M^3 m^4 / (j\hbar)^4$.

An interesting observation related to the equivalence principle can be made that the acceleration, the orbital radius, and orbital velocity are all functions of the mass m in all dimensions as a result of quantization, even though M >> m. The presence of m is not an artifact of the Bohr-Sommerfeld condition. The same mass dependency and basically the same results are obtained from the Schroedinger equation, though by the more difficult route of solving this second order differential equation with associated Laguerre polynomials. The failure of m to vanish indicates that quantum mechanics is inconsistent with the weak equivalence principle (Rabinowitz, 1990, 2001a). This may be related to difficulties encountered in developing a theory of quantum gravity (cf. Sec.10). Classically for M >> m, these variables are independent of the orbiting mass, since m cancels out in accord with the equivalence principle. The equivalence principle requires that the equations of motion be independent of m.

In n-space for M >> m, the total energy of a gravitationally bound atom is

$$E_n = \left[ \frac{(n-4)}{(n-2)} \right] \left[ \frac{G_n M m \Gamma\left(\frac{n}{2}\right)}{\pi^{(n-2)/2}} \right] \left[ \frac{2G_n M m^2 \Gamma\left(\frac{n}{2}\right)}{(j\hbar)^{2(n-2)/(4-n)} \pi^{(n-2)^2/2(4-n)}} \right]^{(n-2)/(4-n)}. \tag{4.8}$$

In 3-space, eq. (4.8) reduces to $E_3 = -\left(m^3/2\right)\left(GM/j\hbar\right)^2$. The binding energy between M and m of such an atom is given by j = 1, and can vary widely from small to exceptionally large depending on M and m. For example let m = $m_{proton}$ = 1.67 x $10^{-27}$ kg. For M = 6.36 x $10^{10}$ kg, with $R_H$ = 9.43 x $10^{-17}$m, we get a binding energy $E_3$ = 3.76 x $10^{12}$ J = 23.5 MeV, with v = 6.72 x $10^7$ m/sec.

Note from equation (4.8) that all energy levels are 0 in 4 and higher dimensional space. As we shall see, this is a general result no matter how strong the gravitational interaction. As shown in Sec. 5, this conclusion would not be changed with consideration of general non-circular orbits. This is also true for electrostatically bound atoms since they have the same dependence on n.

Mathematically this results from the leading factor [(n-4)/(n-2)] in the complicated quantized equation (4.8). Why n 4 leads to $E_n$ 0, can be understood in simpler terms. For a long-range attractive force like gravity with M >> m

$$F_n = \frac{-2\pi G_n Mm\Gamma(n/2)}{\pi^{n/2} r_n^{n-1}} = \frac{-mv_n^2}{r_n} \Rightarrow \frac{1}{2}mv_n^2 = \frac{\pi G_n Mm\Gamma(n/2)}{\pi^{n/2} r_n^{n-2}}, \quad (4.9)$$

where the (n-1) exponent of r in $F_n$ results from Gauss' law in n-space, e.g. $F_3 = -GMm/r^2$ because the area of a sphere $\propto r^2$, since we live in a 3-dimensional macroscopic space. Substituting equation (4.9) into the equation for total energy

$$E_n = \frac{1}{2}mv_n^2 + \frac{-2\pi G_n Mm\Gamma(n/2)}{(n-2)\pi^{n/2} r_n^{n-2}} = \frac{\pi G_n Mm\Gamma(n/2)}{\pi^{n/2} r_n^{n-2}} + \frac{-2\pi G_n Mm\Gamma(n/2)}{(n-2)\pi^{n/2} r_n^{n-2}}$$

$$= \left[\frac{n-4}{n-2}\right] \frac{\pi G_n Mm\Gamma(n/2)}{2\pi^{n/2} r_n^{n-2}} \geq 0 \text{ for } n \geq 4. \quad (4.10)$$

This result, with the same prefactor [(n-4)/(n-2)], applies both classically and quantum mechanically since quantization does not change the sign of the co-factor $\frac{\pi G_n Mm\Gamma(n/2)}{\pi^{n/2} r_n^{n-2}}$ >0, for positive masses or if both masses are negative.

The emitted frequency in n-space for closest allowed energy transitions is

$$\nu_n = \frac{\Delta E}{2\pi\hbar}$$

(4.11)

$$= \frac{(n-4)}{(n-2)2\pi\hbar} \left[\frac{G_n Mm\Gamma\left(\frac{n}{2}\right)}{\pi^{(n-2)/2}}\right] \left[\frac{\left[2G_n Mm^2\Gamma\left(\frac{n}{2}\right)\right]^{(n-2)/(4-n)}}{\hbar^{2(n-2)/(4-n)} \pi^{(n-2)^2/2(4-n)}}\right] \left[\frac{1}{(j+2)^{2(n-2)/(4-n)}} - \frac{1}{j^{2(n-2)/(4-n)}}\right]$$

The principal quantum number changes by an even integer (2, 4, 6, ...) to conserve angular momentum for the system of atom and emitted graviton of spin 2.

In 3-space, equation (4.11) gives the frequency of radiated gravitons

$$\nu_3 = \frac{G_3^2 M^2 m^3}{4\pi\hbar^3} \left|\frac{1}{j^2} - \frac{1}{(j+2)^2}\right| \text{ for } M >> m. \quad (4.12)$$

This discrete gravitational spectrum is identical in form to the electromagnetic spectrum of the hydrogen atom.

Work is underway to detect gravitational waves from distant sources such as neutron stars, binary pulsars, and coalescing black holes. Signals from such sources are expected to have frequencies in the range from 10 Hz to $10^4$ Hz. Discrete gravitational radiation from orbital de-excitation of an ordinary small mass orbiting a LBH can have frequencies in this range as given by eq. (4.12). Such signals, even though small, from potentially nearby sources can compete or interfere with initially large signals from distant sources (Rabinowitz, 2001a).

Note that in 4-space there is no energy spectrum, nor would there be for any other long-range force like the electrostatic force. The results here indicate Euclidean 4-space is singular in that $r_4$ is infinite, and though angular momentum, L, remains finite, L can't be quantized in the usual way because its dependence on $r_4$ vanishes. This will be explored in Sec. 6.

## 5 Higher Dimensional Non-Circular Orbits

In higher dimensional space, central force trajectories are generally neither circular, nor elliptical, as the orbits become non-closed curves. Although only circular orbits have been considered so far, the more complicated central force problem where there is also a radial velocity, yields the same conclusion. Let us take into consideration the effective potential energy. The general case can be put in the form of a one-dimensional radial problem in terms of the effective potential energy of the system,

$$V_n' = V_n + L^2 / 2mr_n^2. \tag{5.1}$$

where $V_n(r)$ is the potential energy of the system, and L is the angular momentum which remains constant because there are no torques in central force motion.

Orbits are not energetically bound when $E_n(r) - V_n'(r_m) \geq 0$, where $r_m$ is the radius of the circular orbit at the maximum of $V_n'$. Eq. (4.8) showed that $E_n$ 0 for quantized orbits. These quantized orbits as well as classical orbits for which $0 \leq E_n < V_n'(r_m)$ are classically, but not quantum mechanically bound. If orbits could be formed in this region, they would be only metastable since the finite width of the potential energy barrier presented by $V_n'$ permits the orbiting body to tunnel out. The general equation of motion that includes radial motion is

$$F_n = \frac{-K_n}{r_n^{n-1}} = m\frac{d^2 r_n}{dt^2} - \frac{L^2}{mr_n^3}. \tag{5.2}$$

For the gravitational force (Rabinowitz, 2001)

$$K_{Gn} = \frac{2\pi G_n Mm \Gamma(n/2)}{\pi^{n/2}}, \tag{5.3}$$

where we will consider the orbiting mass m << M. For the electrostatic force

$$K_{En} = \frac{2\pi R_{En} Qq\Gamma(n/2)}{4\pi\varepsilon\pi^{n/2}},\tag{5.4}$$

where a body of mass m with negative charge q orbits around a positive charge Q. $R_{En}$ is a model dependent factor that relates the electrical force in n-space to the electrical force in 3-space, and $\varepsilon$ is the permittivity of free space.

Let us substitute the potential energy into eq. (5.1) to determine if there is an n that satisfies:

$$E_n(r_m) - V_n'(r_m) = E_n + \frac{K_n}{(n-2)r_m^{n-2}} - \frac{L^2}{2mr_m^2} \geq 0.\tag{5.5}$$

The maximum value of $V_n'$ occurs at $r_m$, and is obtained by setting $dV_n'/dr = 0$. This is the same as dropping the radial force term $md^2r/dt^2$ in the force eq. (5.2):

$$\frac{-K_n}{r_m^{n-1}} = \frac{-L^2}{mr_m^3} \Rightarrow r_m = \left[\frac{mK_n}{L^2}\right]^{1/(n-4)}.\tag{5.6}$$

This is the radius $r_m$ for a circular orbit at the maximum value of $V_n'$. Trajectories with $r > r_m$ are unbound both classically and quantum mechanically. Substituting for $E_n$ from eq. (4.10) into eq. (5.5),

$$\left[\frac{n-4}{n-2}\right]\frac{K_n}{2r_m^{n-2}} + \frac{K_n}{(n-2)r_m^{n-2}} - \frac{L^2}{2mr_m^2} \quad 0.\tag{5.7}$$

Combining the first two terms, and substituting eq.(5.6) into eq. (5.7):

$$\frac{K_n}{2r_m^{n-2}} \geq \frac{L^2}{2m}r_m^{-2} \Rightarrow 1 \geq \frac{L^2}{2mK_n}\left(r_m^{n-4}\right) = \frac{L^2}{2mK_n}\left(\left[\frac{mK_n}{L^2}\right]^{1/(n-4)}\right)^{n-4} = 1.\tag{5.8}$$

Eq. (5.8) implies that the circular orbit at $r = r_m$ is at the highest energy state, and thus

$$E_n(r_m) = V_n'(r_m) > E_n(r_n)\tag{5.9}$$

Let us first look at $E_n$ by means of the uncertainty principle with $p \sim \Delta p \sim \hbar/2\Delta x$, and $r \sim \Delta x$:

$$E_n \sim \frac{(\Delta p)^2}{2m} - \frac{K_n}{(n-2)(\Delta x)^{n-2}} = \frac{\hbar^2}{8mr^2} - \frac{K_n}{(n-2)r^{n-2}}.\tag{5.10}$$

Classical orbits can exist in the region $0 < E_n < V_n'(r_m)$. However, since they would be subject to quantum tunneling, classical orbits would only be metastable. For n  4 and r small enough to make $E_n < 0$, the orbiting body would spiral in to r = 0 both quantum mechanically and classically since then the negative potential energy term dominates in eq. (5.10).

Except for the s = 0 state, identically the same results in 3-space are obtained for the Bohr-Sommerfeld semi-classical approach for gravitationally or

-11-

electrostatically bound atoms as from the Schroedinger equation. Though the latter is done by the more difficult route of solving this second order differential equation with associated Laguerre polynomials. Therefore it is reasonable to expect the same results in higher dimensions. Even if they were to differ, there is no question that the orbiting mass could tunnel out of the finite width effective potential energy barrier.

# 6 Quantization of Angular Momentum in 4-Space

In all dimensions except in 4-space, the dependence of angular momentum, L, on $r_n$ allows the orbital radius to adjust in the quantization of L. This and no binding energy for atoms for 4-space has consequences for the 4-space Kaluza-Klein unification of general relativity and electromagnetism, as well as for string theory. Let us examine the ramifications of the quantization of L, without quantization of r, in 4-space for gravitational and electrostatic atoms

Equating the gravitational force eq. (4.3) to the centripetal force in 4-space for circular orbits of a two-body gravitationally bound atom of reduced mass $\mu = mM/(m+M)$:

$$F_{Gn} = \frac{-2\pi G_n Mm\Gamma(n/2)}{\pi^{n/2} r_n^{n-1}} \xrightarrow{n=4} \frac{-2\pi G_4 Mm}{\pi^2 r_4^3} = \frac{-2R_G G_3 Mm}{\pi r_4^3} = -\mu \frac{v_4^2}{r_4}, \quad (6.1)$$

where $G \equiv G_3 = G_4/R_G$. $R_G$ is a model dependent factor that relates the gravitational force in n-space to the gravitational force in 3-space. Similarly for $R_E$ and the electrical force. It is assumed here that $R_G = R_E$. Solving eq. (6.1) for the angular momentum, $L_G$, of the two-body gravitational atom, and quantizing $L_G$:

$$L_G = \mu v_4 r_4 = [2\mu R_G G_3 Mm/\pi]^{1/2} = j\hbar, \quad (6.2)$$

Equating the electrostatic force (mks units) to the centripetal force in 4-space for a two-body electrostatically bound atom:

$$F_{En} = \frac{2\pi R_{En} Q(-q)\Gamma(n/2)}{4\pi\varepsilon \pi^{n/2} r_n^{n-1}} \xrightarrow{n=4} -\left(\frac{e^2}{4\pi\varepsilon}\right)\frac{2R_E}{\pi r_4^3} = -(\alpha\hbar c)\frac{2R_E}{\pi r_4^3} = -\mu\frac{v_4^2}{r_4}, \quad (6.3)$$

where $\varepsilon$ is the permittivity of free space, $\alpha$ 1/137 is the fine structure constant, and the electronic charge $e = Q = q$. Solving eq. (6.3) for the angular momentum, $L_E$, of the two-body electrostatic atom, and quantizing $L_E$:

$$L_E = \mu v_4 r_4 = [2\mu\alpha\hbar c R_E / \pi]^{1/2} = j\hbar. \tag{6.4}$$

Quantization lets us set $L_E = L_G$, since they are both $= j\hbar$. This yields a condition on the product of the two masses in terms of the Planck mass $M_P$,

$$Mm = \alpha\left(\frac{\hbar c}{G}\right) = \alpha M_P^2. \tag{6.5}$$

This says that the gravitational angular momentum in 4-space can only be quantized if the product of the two masses $Mm = \alpha M_P^2 \approx M_P^2/137$. Empirically, the electron mass can be related to $\alpha$ and the proton mass,

$$m_e = 10.22\alpha^2 M_p \Rightarrow M_p m_e = 10.22\alpha^2 M_p^2. \tag{6.6}$$

It is an interesting coincidence that not only does $\alpha$ enter into eqs. (6.5) and (6.6), but that they can be put into a somewhat similar form, where an extra factor of $10\alpha$ takes us from the macroscopic to the subatomic domain.

There are not enough constraints to determine the factor, $R_{G,E}$, uniquely. If we combine eqs. (6.5) and (6.2) with $m = M$ and $j = 1$,

$$R_G = \frac{\pi\hbar^2}{G\alpha^{3/2}M_P^3}. \tag{6.7}$$

If we combine eqs. (6.5) and (6.4) with $m = M$ and $j = 1$,

$$R_E = \frac{\pi\hbar}{c\alpha^{3/2}M_P}. \tag{6.8}$$

Eqs. (6.8) and (6.7) are equivalent, and both yield $R_{G,E} \approx 8 \times 10^{-32}$ m. $R_{G,E}$ is about 4940 times bigger than the Planck length of $1.62 \times 10^{-35}$ m. Since this is the smallest $R_{G,E}$ that this approach yields, it may be indicative that for string theory the size of the compact dimensions $> \sim 10^{-31}$ m rather than $10^{-35}$ m. Although the energy scale

$$E \approx M_P c^2 \sqrt{\alpha} = \frac{1.2 \times 10^{19} \text{GeV}}{\sqrt{137}} = 10^{18} \text{GeV} \tag{6.9}$$

is an order of magnitude below the Planck energy, it is still quite high for experimental testing.

Eqs. (6.2) and (6.4) imply quantization of products and sums of the masses if $R_G$ and $R_E$ are not quantized. Eq. (6.2) implies

$$\frac{(Mm)^2}{M+m} \propto (j\hbar)^2. \tag{6.10}$$

Eq. (6.4) implies

$$\frac{Mm}{M+m} \propto (j\hbar)^2. \tag{6.11}$$

# 7 Accelerated Expansion Of The Universe

Little black holes are excellent candidates for the missing mass of the universe, which is up to 95% of the $10^{53}$ kg mass of the universe. Although they are quite hot, they qualify as cold dark matter since their velocities << c. One piece of evidence for dark matter comes from spiral galaxies, as first observed by Vera Rubin (1983). The missing mass gives the stars ~ constant linear velocities around their galactic center independent of radial distance r, rather than the expected Keplerian velocities $\propto 1/\sqrt{r}$. Other evidence comes from clusters of galaxies. The speeds at which individual galaxies are moving in these clusters are so high that the clusters would fly apart unless they were held together by a stronger gravitational attraction than provided by the masses of all the galaxies.

The little black hole GTR model discussed in Section 2 is much less likely to interfere with nucleosynthesis than Hawking's because the radiation is much reduced, and is beamed rather than omnidirectional. The radiation from Hawking little black holes would likely have either interfered with early universe nucleosynthesis or broken up products of nucleosynthesis after the nuclear reactions were over. Furthermore, that many Hawking LBH would fry the universe.

The quiescent LBH of this paper can be much smaller than previously considered since the GTR power is proportional to the very small transmission coefficient. The smallest mass that can survive within the age of our universe, ~ $10^{17}$ sec is (Rabinowitz, 1999 a, c)

$$M_{small} \geq 10^{12} \left\langle e^{-2\Delta\gamma} \right\rangle^{1/3} \text{ kg} . \tag{7.1}$$

The Hawking result of $10^{12}$ kg is obtained by setting $e^{-2\Delta\gamma}$ = 1 in eq. (7.1). We should bear in mind that even with Hawking radiation, there may be large numbers of LBH that have more recently decayed to mass well below $10^{12}$ kg. Since 0 $e^{-2\Delta\gamma}$ 1, an entire range of primordial black hole masses << $10^{12}$ kg may have survived from the beginning of the universe to the present than permitted by Hawking's theory.

For example, if the average tunneling probability $\left\langle e^{-2\Delta\gamma} \right\rangle$ ~ $10^{-45}$, then $M_{small}$ ~ $10^{-3}$ kg. For $M_{univ}$ ~ $10^{53}$ kg, $V_{univ}$ ~ $10^{79}$ m$^3$ (radius of 15 x$10^9$ light-year = 1.4 x $10^{26}$ m), the average density of such LBH would be 1 LBH per $10^{23}$m$^3$ . The velocity of our local group of galaxies with respect to the microwave background (cosmic rest frame), $v_{LBH}$ ~ 6.2 x $10^5$ m/sec (Turner and Tyson, 1999), is a reasonable velocity for LBH with respect to the Earth. This may make it possible to detect their incident flux ~ ($10^{-23}$/m$^3$)(6.2 x $10^5$ m/sec) ~ $10^{-17}$/m$^2$sec on the Earth. (Rabinowitz, 2001a)

Primordial GTR black holes would be much less subject to the limits imposed by nucleosynthesis arguments than the Hawking model. The baryonic matter in them would have bypassed the deuterium and helium formation that occurred during the era of nucleosynthesis. That stars orbit with ~ constant linear velocity may result from a linear increase in total mass of the ensemble of little black holes with radial distance from a galactic center. This is due to radiation reaction force driving them outward as well as a lower evaporation rate at larger radial distances. The analysis for such galaxies would be similar to that which will be done in the latter part of this section for the universe.

This mechanism may also be able to account, at least in part, for the recently observed accelerated expansion of the universe, as shown in the analysis below. One of the most remarkable and exciting discoveries of 1998 revealed that the universe is accelerating in its expansion (Perlmutter, et al, 1998) and (Riess et al ,1998). The implication is that the universe is older, bigger, and less dense than previously thought. This discovery calls into question long-standing cosmological theories. It may shed light on the enigma that some stars appear to be older than the previously accepted age of the universe.

If LBH represent a substantial fraction (up to 95%, because in terms of "smoothness" they can be mistaken for energy, since they are so small they can be "smoothed over") of the missing mass of the universe, then their radially inward directed radiation in interacting with the universe as the second body is a good candidate model for the accelerated expansion of the universe. Einstein's cosmological constant and inflation have been considered as explanations. However a big cosmological constant makes the vacuum enormously more massive than is consistent with observation or quantum theory, as will be discussed in Sec. 9. Directed radiation from LBH is a possible explanation that does not have these problems.

Let us determine the maximum acceleration observable from the Earth due to the radiation reaction force experienced by a LBH of mass M in a spherical shell at radius $R_U$, surrounding mass $M_U$ of the universe. For a first approximation, let us ignore the deceleration caused by gravity. Bodies at the edge of the universe are moving out radially near the speed of light, thus requiring relativistic treatment. As seen from the Earth, the acceleration is

$$a \equiv \frac{dv}{dt} < \frac{-c}{M}(1-\beta^2)\frac{dM}{dt}, \quad (7.2)$$

where $\beta \equiv v/c$, and $\frac{dM}{dt} = \frac{-P_R}{c^2}$. $P_R \approx \left|\frac{\hbar c^6 \langle e^{-2\Delta\gamma}\rangle}{16\pi G^2}\right|\frac{1}{M^2}$ as given by eq. (2.1).

The < sign applies because gravitational attraction was neglected. Solving Schroedinger's equation for $\Delta\gamma$ (Rabinowitz, 1999 a, c):

$$\Delta\gamma \approx \frac{m_{em}}{\hbar}\cdot\sqrt{\frac{2GM}{d}}\left\{\sqrt{b_2(b_2-d)}-\sqrt{b_1(b_1-d)}-d\,\ell n\left|\frac{\sqrt{b_2}+\sqrt{b_2-d}}{\sqrt{b_1}+\sqrt{b_1-d}}\right|\right\}, \quad (7.3)$$

where $m_{em}$ is the mass of an emitted particle. The solution applies for $R \gg b_2$, which is valid for $M_U = M_2 \gg M$, since $b_2$ and $b_1$ are both $\gg d$; and

-15-

$$d = Mb_1(R-b_1)R/[M(R-b_1)R+M_2(b_1)^2] \approx MR_U/M_U. \qquad (7.4)$$

$$m_{em} \approx \frac{k[T]}{c^2} = \frac{k}{c^2}\left[\frac{\hbar c^3}{4\pi kGM}\right] = \left[\frac{\hbar c}{4\pi GM}\right], \qquad (7.5)$$

where T was obtained from eq. (3.1). Substituting eqs. (7.4) and (7.5) into eq. (7.3)

$$\Delta\gamma \approx \left(\frac{m}{\hbar}\right)\sqrt{\frac{2GM}{d}}\{b_2 - b_1\} = \left(\frac{c}{4\pi GM}\right)\sqrt{\frac{2GM}{(MR_U/M_U)}}\{b_2 - b_1\} \qquad (7.6)$$

Substituting eqs. (2.1) and (7.6) into eq. (7.2):

$$a = \frac{(1-\beta^2)}{M^3}\left[\frac{\hbar c^5}{16\pi G^2}\right]\left\langle\exp-\left[c\frac{b_2-b_1}{2\pi M}\sqrt{\frac{2M_U}{GR_U}}\right]\right\rangle. \qquad (7.7)$$

Taking $da/dM = 0$, in eq.(7.7) we can find the relationship between M and $(b_2 - b_1)$ for maximum acceleration,

$$\frac{(b_2-b_1)}{M} = \left[\frac{c}{6\pi}\sqrt{\frac{2M_U}{GR_U}}\right]^{-1}. \qquad (7.8)$$

Substituting eq. (7.8) into eq. (7.9) greatly simplifies the exponential, giving the maximum possible acceleration:

$$a_{max} \leq \frac{(1-\beta^2)}{M^3}\left[\frac{\hbar c^5}{16\pi G^2}\right]e^{-3} = 5.7 \times 10^{25}\frac{(1-\beta^2)}{M_{kg}^3} \, m/\sec^2. \qquad (7.9)$$

Note that eq. (7.9) is quite general as it does not depend on the particular form of the exponent, $\Delta\gamma$, but only on the dependence of $\Delta\gamma$ on M.

Strictly speaking, a spherically symmetric universe is only rigorously homogeneous with respect to its center. However, if it is infinite, this would not show up in experimental measurements such as the uniformity in all directions of the microwave background radiation. If the universe is finite, measurements made at distances relatively close to its center (compared with its radial size) would tend to hardly show the inhomogeneity. In this case, the inhomogeneity of off-center observations would tend to be lost in the experimental uncertainties. LBH would also yield accelerated expansion for a non-Euclidean universe since each LBH would be repelled from its neighbors by the beamed radiation. Expansion in all geometries gives the appearance that any point may be considered to be central as all points appear to be expanding away from all others.

For a numerical example, if the average LBH mass in a spherical shell is ~ $10^9$ kg, then $a_{max} <\sim 10^{-1}$ m/sec$^2$. Although the accelerated expansion of the universe is well established (Perlmutter et al, 1998 and Riess et al, 1998), to date no experimental value of the acceleration has been determined.

If the total mass of a spherical shell of the universe is dominated by an ensemble of little black holes, then their acceleration will transport the rest of the bodies in the shell with them by gravitational attraction.  The acceleration of each shell will not exceed the value given by eq. (7.9).  Interestingly, even though $R_U$, the radius of the universe, was used to derive eq. (7.9), it is independent of $R_U$. Correction for the deceleration caused by gravity does depend on radial distance.

      The momentum transfer of beamed GTR can either be the main mechanism or just contribute to the accelerated expansion of the universe.  As the universe gets old and the LBH radiate away to nothing,  GTR predicts that the accelerated expansion of the universe will decrease and eventually the acceleration will cease. This is in contrast to the exponential accelerated expansion of the universe that may be obtained from a positive cosmological constant, $\Lambda$, of Einstein's general relativity.  A positive $\Lambda$ predicts that the universe will endlessly expand quicker and quicker.  GTR's prediction is also in contrast to models that predict a period of accelerated expansion followed by contraction.

## 8  GTR Avoids the Black Hole Lost Information Paradox

      Hawking concludes that the information that entered a black hole can be forever lost (Hawking, 1976).  This comes directly from his assumption that black hole radiation is black body radiation, which loses information about its source. It is not necessary to attribute this to his position that the radiation does not originate from within a black hole, but comes from the vicinity outside it due to particle anti-particle creation. If it only appears to come from within, then it would not  reflect what is inside as the hole evaporates away.  He argued that as the black hole radiates, it will eventually completely evaporate away.  The resulting radiation state would be precisely thermal so there would be no way to retrieve the initial state.

      On the other hand, time-reversal symmetry, classical and quantum physics are violated if the information is lost. A consequence of  gravitational tunneling radiation (GTR) is that it resolves this enigma, since it comes from within the hole and carries attenuated but undistorted information from within. Since it is a tunneling process and not an information voiding Planckian black body radiation distribution, it can carry information related to the formation of a black hole, and avoid the information paradox associated with Hawking radiation.

## 9  Vacuum Energy and the  Cosmological Constant

Aristotle wrote that nature abhors a vacuum, and the zero-point energy of modern physics seconds this notion in which a vacuum abounds with field energy fluctuations, and with particle-antiparticle pairs that spontaneously pop in and out of existence.  Surprisingly, it leads to detectable effects like the Lamb shift of atomic energy levels and the Casimir effect that are found to agree with theory to high accuracy.  But this comes with a high price to pay.  Though most of the effects are finite, the calculated total zero-point energy is infinite over any finite volume of space.

This infinity can be reduced to a finite energy by somewhat arbitrary theoretical cut-offs, such as not allowing wavelengths shorter than the Planck length ($10^{-33}$ cm).  This is inconsistent with special relativity, as wavelength is a function of the observer's frame of reference.  Nevertheless, it is done -- but only with hollow success.  Another approach relates to the standard model of fundamental particles.  In these approaches, the resulting energy is finite, but is between $10^{46}$ to $10^{120}$ times too large.  This divergence between experience and theory is perhaps the most bewildering problem in physics, whose resolution may lead to far-reaching consequences of our view of nature.

The cosmological constant, $\Lambda$, of general relativity is proportional to the energy density of the vacuum.  $\Lambda$ may be negative, zero, or positive.  With the above exceptionally large vacuum energy, depending on the sign of $\Lambda$, according to general relativity the universe would have to be much smaller ($-\Lambda$) or larger ($+\Lambda$) than it is.  There is a problem even without $\Lambda$, as such an enormous vacuum energy would give the universe a strong positive curvature -- contrary to observations of a flat universe.  An expected consequence of GTR is that it can allow for a smaller value of the cosmological constant, if GTR contributes to the accelerated expansion of the universe.

 When all matter and heat radiation have been removed from a region of space, even classically there remains a pattern of zero-point electromagnetic field energy throughout the space.  Nevertheless, as an interesting exercise let us see how much of the discrepancy can be reduced if we assume that the vacuum zero-point energy is only associated with matter, i.e. it only exists in the vicinity of matter due to a rapid attenuation of fields (Rabinowitz, 2001b).  In contrast to the prevailing view,  in my speculative hypothesis this energy would be limited to the region of space near matter and there would only be negligible vacuum zero-point energy in the vast regions of space where there is no matter. If the volume of matter in the universe is mainly nucleons (5% of the mass of the universe), we can easily estimate the reduction:

$$\frac{V_{Universe}}{V_{TotalNucleons}} \sim \frac{10^{79} \text{m}^3}{\left(\frac{.05 M_{Universe}}{M_{Nucleon}}\right) V_{Nucleon}} \sim \frac{10^{79} \text{m}^3}{\left(\frac{.05 \times 10^{53} \text{kg}}{10^{-27} \text{kg}}\right) 10^{-45} \text{m}^3} = 2 \times 10^{45} \quad (9.1)$$

Thus the vacuum zero-point energy can be reduced to between $10^{46-45} = 10^1$ and $10^{120-45} = 10^{75}$ times too large if this premise has merit.

-18-

The needed $10^{120}$ may be achieved with a universe filled with 95% LBH of average mass $10^{-7}$ kg ($10^{-4}$ gm)

$$\frac{V_{Universe}}{V_{TotalLBH}} = \frac{V_{Universe}}{\left(\frac{0.95 M_{Univ}}{M_{LBH}}\right) V_{LBH}} \sim \frac{10^{79} m^3}{\left(\frac{10^{53} kg}{10^{-7} kg}\right)\left[\frac{4}{3}\pi(1.5 \times 10^{-34} m)^3\right]}$$

$$= \frac{\rho_{LBH}}{\rho_{Universe}} \sim \frac{10^{91} gm/cm^3}{10^{-29} gm/cm^3} = 10^{120} \qquad . \qquad (9.2)$$

## 10 Quantum Mechanics Violates the Equivalence Principle

As shown in Sec. 4, the possible replacement of Hawking radiation with gravitational tunneling radiation allows the existence of gravitationally bound atoms. Analysis of such atoms led to a demonstration that quantum mechanics is inconsistent with the weak equivalence principle (EP) because the orbiting mass m remains in the quantized equations of motion, even for M >> m. In turn this raises the question of whether it will be possible to combine Einstein's general relativity (EGR) and quantum mechanics into a theory of quantum gravity. So far attempts over many decades have all led to discrepancies and even contradictions. A theory of quantum gravity, QG, may have far-reaching astrophysical implications. QG could shed light on the big bang and the early universe. QG may also determine fundamental physical attributes that start on a small scale and affect large scale astrophysical properties. So it is important to explore this potential incompatibility. First let us clearly state what EP is.

In EGR, test body motion is along geodesics that are undisturbed by the test body. A geodesic (the shortest distance between two points in a given space) is a geometrical property independent of the physical characteristics of the test body as long as it produces a negligible perturbation of the geodesic. Not all authorities agree on what the weak equivalence principle (WEP) is, and what the the strong equivalence principle (SEP) is of EGR. For example, one view of WEP is: 1) that there are no physical effects that depend on the mass of a point particle in an external gravitational field. This is related to the equivalence of inertial and gravitational mass. For those with this interpretation, SEP asserts: 2) that being at rest in a gravitational field is locally equivalent to being at rest in an accelerated system. Clifford Will (Davies, 1989) essentially accepts 1) for WEP and calls a more rigorous equivalent of 2) the Einstein equivalence principle (EEP): 3) (i) WEP is valid; (ii) results of local non-gravitational experiments are independent of the velocity of their freely falling reference frames; (iii) results of local non-gravitational experiments are independent of where and when they are performed.

Will goes on to formulate an even stronger principle than EEP, which he calls SEP: 4) (i) all bodies (miniscule to ultra-stellar) with their own internal binding energy (gravitational, etc.), fall with the same acceleration in an external gravitational field; (ii) even in a large falling frame ( e.g. star), the laws of gravity as well as the laws of non-gravitational physics are independent of the velocity and location of the frame relative to other matter.  Rohrlich (1965) doesn't use the terms WEP and SEP.  Will's 4) is similar to Rohrlich's general statement of EP:  5) The equations of motion of a nonrotating test body in free fall in a gravitational field are independent of the energy content of that body.  He appears to diminish the role of SEP in saying (p.42): "True gravitational fields can never be transformed away. ... Apparent gravitational fields are a characteristic of the motion of the observer (rather than of the observed physical system), while true gravitational fields  are the same for all observers no matter what their motion. ... Einstein's statement [of EP] as an equivalence between accelerated observers and gravitational fields is now seen to be restricted to apparent gravitational fields.  True gavitational fields cannot be simulated by acceleration (i.e. by a coordinate transformation)."

The different statements of the equivalence principle are interrelated. If the equations of motion involve the mass, m, it appears that  not only is WEP violated but this also involves a violation of SEP.  (One would expect m to cancel out when averaging over states with large quantum numbers that puts them effectively in the classical continuum.) In quantum mechanics, the wavelength is inversely proportional to the momentum and hence involves the mass.  Quantum mechanical interference effects in general, and quantum-gravitational interference effects in particular, depend on the phase which depends on the mass. This is intrinsic to quantum mechanics.  The weak equivalence principle requires that the equations of motion be independent of m.  Even though SEP works well independently of QM, this is not the case in the union of general relativity and quantum mechanics (quantum gravity).

It is clear that SEP $\Rightarrow$ WEP since SEP yields equations of motion of a body independent of its mass in a gravitational field.   In logic, if A $\Rightarrow$ B then not B $\Rightarrow$ not A. That is, not WEP $\Rightarrow$ not SEP.  Quantum mechanics clearly violates WEP and logic shows that a violation of WEP implies a violation of SEP. Therefore the quest for a theory of quantum gravity seems doomed, unless it is not based upon the equivalence principle, or if quantum mechanics can eliminate its mass dependence.  Neither of these seem likely at the present time.

## 11 Discussion

Although general relativity deals with geodesics in curved space-time, one may accurately depict Einsteinian black holes as having an effective potential energy $\propto 1/r$, the same as Newtonian black holes at $r > 10$ $R_H$. The differences between the two are only near the black hole and inside. Concerns related to tunneling out of Einsteinian black holes may be avoided by considering the black holes here to be Newtonian. Einstein's general relativity (EGR) becomes very non-linear at a black hole and breaks down at the singularity in the center. It may not even give a correct representation of a black hole. Einstein himself was troubled with the nature of black holes in general relativity. General relativity fails at very small distances, as do Newton's and Coulomb's laws due to quantum effects. At present there is no direct or indirect experimental evidence concerning the regions very near or inside a black hole. The model derived here allows for indirect testing of the nature of black holes through the observed radiation.

The concept of black holes whose gravitational attraction is so great that even light may not be expected to escape, can be traced back to John Michell in 1783. A little over a century later, on Nov. 25, 1915, Albert Einstein published a short communication version of his Theory of General Relativity, one of the most profound theories conceived by the human mind. Less than two months later, in 1916, Karl Schwarzschild derived a spherically symmetric solution of Einstein's general relativity for an uncharged black hole of mass M with no angular momentum, which he sent to Einstein for submission to a journal or the Academy (Chandrasekhar, 1983) It is remarkable that not only did Schwarzschild do this so quickly, but that he did it while fatally ill with pemphigus (a debilitating disease with burning blisters on the skin and mucous membranes), while in active combat for the German Army on the Russian Front in WWI. Einstein presented Schwarzschild's solution to the Prussian Academy, on his behalf on Jan. 13, 1916. Unfortunately, Schwarzschild died June 19, 1916 -- too soon to be aware of the importance of his work.

The Schwarzschild metric is

$$ds^2 = -[1 - \frac{2GM}{c^2 r}]c^2 dt^2 + [1 - \frac{2GM}{c^2 r}]^{-1} dr^2 - r^2 [d\Theta^2 + \sin^2 \Theta d\Phi^2], \qquad (11.1)$$

where the units for G and c are usually set = 1. Since the metric coefficients are explicitly independent of time and there is no dragging of the inertial frame, the space-time is static for an observer outside the black hole and t is the proper time if the observer is at rest at infinity. The singularity at r = 0 (the center of the hole) is unavoidable within Einstein's theory. The time and space singularities at the Schwarzschild (horizon) radius $R_H = (2GM/c^2)$ are artifacts, and can be removed by a change of coordinates. A full understanding of Schwarzschild space-time was achieved only in the last several decades (Finkelstein, 1958).

In EGR all fields, except the gravitational field, produce space-time curvature, since the gravitational field is but a consequence of the curvature of space-time. Einstein reasoned that since curvature produced gravity, curvature cannot change gravity, *i.e.* make more or less gravity. To Einstein this would be double counting. The argument becomes less clear when inverted: Gravity cannot change curvature. This prohibition in EGR ultimately leads to black holes, in which singularities are the most egregious difficulties.  Furthermore, Weinberg (1972) has shown that most of the features of the gravitational field can be derived from its symmetry properties as is the case for all other fields in quantum theory.  This tends to support the view that gravity may not be intrinsically due to space-time curvature.

This debate has not yet been decided, and there is even a laboratory experiment that seems to favor a non space-time curvature explanaton of gravity because of its violation of the weak equivalence principle.  Until this experiment was done, it was not clear that even weak-field gravity could be incorporated into quantum mechanics (Overhauser and Colella, 1974).  Colella, Overhauser, and Werner (1975) performed a remarkable experiment showing that quantum mechanics could incorporate the gravitational field and describe its effect on neutron diffraction despite being inconsistent with WEP.  This is reviewed and closely analyzed by Greenberger and Overhauser(1979).

Hüseyin Yilmaz (1958) developed an interesting variation of EGR which avoids the black holes altogether.  He assumed that gravitational field energy also produces curvature of space-time by adding a "gravitational stress-energy tensor" to Einstein's equations.  Since this term is relatively small in the three major tests of EGR, Yilmaz' general relativity (YGR) makes essentially the same predictions as EGR for the advance of the perihelion of mercury, gravitational red shift, and the bending of starlight (the least accurately measured of the three tests).

It may not be obvious why including gravitational field energy as a source of space-time curvature eliminates black holes.  Here is a simple intuitive way to understand this.  The static field and the near field (induction field) of a time-varying gravitational field have negative energy.  (This can be tricky since the electrostatic field energy is positive.  The radiation field has positive energy for both fields.)  Negative energy gives negative curvature tending to cancel the positive curvature due to mass.  Instead of black holes, YGR has grey holes where the emitted light is greatly red-shifted.  If neutron stars with mass much greater than 3 solar masses, or white dwarfs with mass much greater than 1.4 solar masses were detected, this would favor YGR over EGR.

Despite much effort, no one has yet solved the two-body problem in EGR. (This is another reason why GTR was not derived in the context of EGR.)  Yilmaz and his colleagues would say that they never will since it is their view that EGR is a one-body theory in which one body (e.g. the sun) establishes a space-time curvature (field) which determines the motion of a test body (e.g. Mercury) that hardly perturbs the established field.  YGR claims to be an N-body theory. Yilmaz says that if one takes the weak field limit of EGR and considers a many body problem like the perturbation effects of planetary orbits on each other, EGR doesn't work. He asserts that the weak field limit of YGR not only gives the same  perturbation effects as Newtonian gravity, but also the correct precession of Mercury.  Another point in favor of Yilmaz is that in particle physics, energy is ascribed to the gravitational field. Honest dissension is a major mode by which physics progresses and comes into closer agreement with nature. Whether EGR or YGR is correct, needs ultimately to be decided by experiment.

Though black holes were long considered to be a fiction, they now appear to be accepted by most astrophysicists (Genzel, 1998). Neither Hawking nor this paper deal with quantum gravity. Hawking basically deals with a single "isolated" black hole. To my knowledge, the general relativistic solution for a black hole in the presence of other bodies has not been derived as yet. Black hole radii may be expected to be not only a function of their own mass, but also a function (in different directions) of the mass of nearby bodies. The two body problem is at the heart of my calculation, as the second body thins the barrier.

To properly derive radiation from a black hole, a theory of quantum gravity is needed. Despite intense theoretical effort, a theory of quantum gravity is still not in sight. In Hawking's model of black hole radiation, quantum field theory is superimposed on curved spacetime, and gravity is described classically according to Einstein's general relativity. It is a semiclassical approach in which only the matter fields are quantized, black hole evaporation is driven by quantum fluctuations of these fields, and infinities such as the infinite radiation frequencies at the black hole horizon are manipulated in ways that do not seem justified.

Gravitational tunneling radiation is also derived semiclassically, so it is subject to similar criticism. However, infinities do not arise and there is no need to manipulate them in deriving GTR. It is clear from the history of physics that the use of rudimentary theories, like the Bohr theory for the hydrogen atom, can lead to accurate results. Furthermore, their simplicity can lead to physical insights that may be missed when turbid advanced theories are applied.

## 12 Conclusion

Occasionally theories become established without a good foundation of supporting experiments. They become part of the lore of physics and are difficult to displace by a competing theory. Until the old theory is proven wrong by experiment, new paradigms are considered wrong by definition since they predict results which differ from those of the established, hence "correct," theory. This has been the case for Hawking's seniority model of black hole radiation, which has been theoretically but not experimentally established. Belinski (1995), a noted authority in the field of general relativity, unequivocally concludes "the effect [Hawking radiation] does not exist."

If the Yilmaz general theory of relativity is correct and there are no black holes, this is devastating for Hawking's model of black hole radiation, but has no serious ramifications for my model of gravitational tunneling radiation.

Tunneling radiation would still be nearly the same between a very dense little grey hole and a second body.

Since Hawking radiation has proven elusive to detect for three decades, a goal of this paper has been to present an alternative model and its consequences, which can be experimentally tested. This paper gives an insight as to why this may be so, as the radiated power from a black hole derived here $P_R \propto e^{-2\Delta\gamma} P_{SH}$ depends strongly on the factor $e^{-2\Delta\gamma}$. In this epoch, for a given mass, the tunneling probability for emitted particles can result in a radiated power many orders of magnitude smaller than that calculated by Hawking. This lower radiated power permits the survival of much smaller primordial black holes from the early black hole-dominated universe to the present.

Little black holes may be able to account for the missing mass of the universe. In a very young compact universe, radially directed tunneling radiation would have been substantial and may have contributed to the expansion of the early universe. In later epochs this radially directed radiation can help clarify the recently discovered accelerated expansion of the universe.

Examination of quantized gravitational orbits demonstrates an inconsistency between quantum mechanics and the weak equivalence principle. Analysis of higher dimensions indicates that angular momentum cannot be quantized in the usual manner in the fourth dimension. This and no binding energy for atoms for 4-space has consequences for Kaluza-Klein theory, as well as for string theory. Exploration of other ways to quantize angular momentum in 4-space leads to quantization of Mm/M+m, etc. It was found that quantized gravitational or electrostatic atoms are not energetically stable in dimensions higher than three. This may impact string theory if short range forces or other constraints cannot be invoked to achieve stability when the extra dimensions are unfurled. A conjectural hypothesis is offered to help reduce the enormously too-high zero-point energy of the vacuum.

## Acknowledgment

I wish to thank Mark Davidson for helpful discussions.

## References

Allen, B. et al (1999) *Phys. Rev. Lett.* **83**, 1498.
Argyres, P.C., Dimopoulos, S., and March-Russell, J. (1998) *Phys.Lett.* **B441**, 96.
Bardeen, J.M.,Carter, B., and Hawking,S.H. (1973) *Commun. Math. Phys.* **31**, 161.
Belinski, V.A. (1995) *Phys.Lett.* **A209**, 13 .
Chandrasekhar, S. (1983) *The Mathematical Theory of Black Holes*, Oxford Univ. Press, New York.
Colella, R., Overhauser, A. W., and Werner, S.A. (1975) *Phys. Rev. Lett.* **34**, 1472.
Davies, P. (1992) *The New Physics*, Chapter 2. Cambridge Univ. Press, Great Britain.

Dunning-Davies, J. and Lavenda, B. H. (1998) *Physics Essays* **3**, 375.
Finkelstein, D. (1958) *Physical Review* **110**, 965.
Genzel, R. *Nature* (1998) **391**, 17.
Greenberger, D.M. and Overhauser, A. W. (1979) *Rev. Mod. Phys.* **51**, 43.
Hawking, S. W. (1974) *Nature* **248**, 30.
Hawking, S. W. (1975) *Commun. Math. Phys.* **43**, 199.
Hawking, S. W. (1976) *Phys. Rev. D* **14**, 2460.
Overhauser, A. W. and Colella, R. (1974) *Phys. Rev. Lett.* **33**, 1237.
Perlmutter, S. et al, (1998) *Nature* **391**, 51.
Rabinowitz, M. (1990) *IEEE Power Engineering Review* **10**, No. 4, 27.
Rabinowitz, M. (1999a) *Astrophysics and Space Science* **262**, 391.
Rabinowitz, M. (1999b) *IEEE Power Engineering Review Letters* **19**, No.3, 65.
Rabinowitz, M. (1999c) *Physics Essays* **12**, 346.
Rabinowitz, M. (2001a) *Int'l Journal of Theoretical Physics*, **40**, 875.
Rabinowitz, M. (2001b) *Journal of New Energy* **6**, 113.
Rohrlich, F. (1965) *Classical Charged Particles*, Addison-Wesley, Reading, MA.
Riess, A.G. et al (1998) *Astronomical Journal* **116**, 1009.
Rubin, V.C. (1983) *Science* **220**, 1339.
Thorne, K.S., Price, R.H., and Macdonald, D. A. (1986). *Black Holes: The Membrane Paradigm*, Yale, New Haven CT.
Turner, M. S. and Tyson, J. A. (1999) *Reviews of Modern Physics* **71**, S145.
Weinberg, S. (1972), *Gravitation and Cosmology*, Wiley, New York.
Yilmaz, Hüseyin (1958) *Phys. Rev.* **111**, 1417.